%% file: final_submit.tex
\documentclass[12pt, draftclsnofoot, onecolumn]{IEEEtran}
\IEEEoverridecommandlockouts

\usepackage{amsmath,amssymb,dsfont,stfloats,color,url}
\usepackage[pdftex]{graphicx}
\usepackage{subfigure}
\usepackage{tikz}
\usepackage{algpseudocode,algorithm,algorithmicx}
\PassOptionsToPackage{draft}{hyperref}
\usepackage{cite}

\usepackage{graphicx,color,epsfig,rotating,subfigure}

\usepackage{balance}
\usepackage{cite}
\usepackage{subfigure}
\usepackage{url}
\usepackage{array}
\usepackage{verbatim}
\usepackage{stfloats}

\usepackage{color}
\definecolor{red}{rgb}{1,0,0}
\definecolor{blue}{rgb}{0,0,1}

\def\argmin{\mathop{\rm arg\,min}}

\usepackage{float}
\usepackage{afterpage}
\input{symbols.tex}

\input{macros.tex}

\makeatletter
\newcommand{\ssymbol}[1]{^{\@fnsymbol{#1}}}
\makeatother

\graphicspath{ {./Figures/} }
\begin{document}

\title{Spatially Scalable Lossy Coded Caching}


\author{\IEEEauthorblockN{Mozhgan~Bayat,~\c{C}a\u{g}kan~Yapar, and~Giuseppe~Caire\IEEEauthorrefmark{1}}\\
	\IEEEauthorblockA{Communications and Information Theory Group, Technische Universit\"{a}t Berlin, 10623 Berlin, Germany}\\
	\IEEEauthorblockA{E-mails: \{bayat, cagkan.yapar, caire\}@tu-berlin.de}
	\thanks{\IEEEauthorrefmark{1}The authors thank Saeid Haghighatshoar for his useful comments. \\This work was funded by the European Research Council  under the ERC Advanced Grant N. 789190, CARENET.}}

\maketitle

\begin{abstract}
We apply the coded caching scheme proposed by Maddah-Ali and Niesen to a multipoint multicasting video paradigm.
Partially caching the video files on the wireless devices provides  an opportunity to decrease data traffic load in peak hours via sending 
multicast coded messages to users. In this paper, we propose  a two-hop wireless network for video multicasting, where
the common coded multicast message is transmitted through different single antenna {\em Edge Nodes} (ENs) to multiple antenna users. 
Each user can decide to decode any EN by using a zero forcing receiver.  Motivated by {\em Scalable Video Coding} (SVC), we consider successive refinement source coding 
in order to provide  a ``softer'' tradeoff between the number of decoded ENs and the source distortion at each user receiver. 
The resulting coding scheme can be seen as the concatenation of Maddah-Ali and Niesen coded caching for each source-coded layer, and 
multiple description coding. Using stochastic geometry,  we investigate the tradeoff between delivery time and per-user 
average source distortion. The proposed system is {\em spatially scalable} in the sense that, for given  users' and ENs'  spatial density, 
the achieved distortion-delivery time performance is independent of the coverage area (for in the limit of large area). 
\end{abstract}

\section{Introduction}
\label{sec:Introduction}
Due to the ever-increasing traffic generated by the wireless Internet and the scarcity of bandwidth, it is imperative to develop multicasting algorithms that exploit the recent improvements in storage capacity and computational power. With video delivery (originating from services such as YouTube, Netflix, Amazon Prime, etc.) being responsible for the majority of IP traffic \cite{index2017zettabyte}, this topic becomes especially relevant. Taking the users' consumption patterns into account plays a significant role in alleviating the burden of increased traffic during busy hours. Caching addresses this issue by duplicating parts of the content in the end user's storage during off-peak hours and using these copies to recreate the content during peak hours, where the user's demands are known. Maddah-Ali and Niesen (MAN) \cite{maddah2014fundamental} proposed a novel coded caching scheme that tackled this issue by optimizing both the placement and delivery phase for different demands, thus ensuring that a global caching gain is achieved. 
There has been an emergence of recent standards that were designed to multicast streaming content (e.g. live TV), such as evolved Multimedia Broadcast Multicast Service (eMBMS). It is possible to extend these paradigms by applying the above-mentioned coded caching scheme to design an on-demand video content delivery service
in a lossy transmission environment.
In such an environment, users experience different quality of video according to the respective channel quality. 
{\em Scalable Video Coding} (SVC) \cite{schwarz2007overview} refers to a set of techniques which allows to a smooth tradeoff between 
video quality and source coding rate and  SVC encoder produces layers of source-encoded bits. Layers are decoded in sequence at each user receiver, 
such that the number of successive decoded layer(s) determines the reconstructed source quality. 

A well-known information theoretic setting was introduced in \cite{equitz1991successive}, known as successive refinement source coding.
In this paper we focus on the special case of two layers.  The encoder provides a fundamental layer at rate $R^S_1$ and a refinement layer at rate 
$R_2^S$. A source is said to be {\em successively refinable} if the points 
$(D_1, R_1^S)$ and $(D_2, R_1^S + R_2^S)$ achievable by successive refinement are points on the boundary of the source distortion-rate region. 
It is well-known that Gaussian sources with Mean Square-Error (MSE) distortion are successively refinable. It is also well-known that successive refinability 
with respect to the MSE distortion holds approximately for many other source distributions. 
In this paper, in order to gain theoretical insights into the problem, we consider the Gaussian source case. For an i.i.d. source $\sim \Nc(0,\sigma^2)$, we have
$D_1 = \sigma^2 2^{-2 R^S_1}$ and $D_2 = \sigma^2 2^{-2(R_1^S + R_2^S)}$. 

Successive refinement has been applied to different  coded caching scenarios.  
The authors in \cite{sengupta2016layered} presented a framework for layered caching in which heterogeneous 
cache sizes are considered for the end users. In \cite{yang2016coded, cao2018coded} a coded caching scheme with fixed heterogeneous distortion requirements is investigated.
They provided a lower bound for the delivery rate in a scenario with two users  and a library consisting of $N>2$ files.
In \cite{hassanzadeh2015distortion}, a cache-aided network for video delivery is considered. 
The delivery scheme was designed to minimize the average per-user distortion. 
During the placement phase, a user partially caches the layers according to popularity distribution of the files. 
In \cite{7541369} the authors considered equal distortion requirements for all users. 
They provided an inner and outer bound for a multi-layer coded caching scheme by treating the cache memory data as side information.
Finally, the authors in \cite{ibrahim2018coded} provided an algorithm to jointly encode subfiles of different layers 
to achieve a lower rate than \cite{yang2016coded}.
The works \cite{yang2016coded,cao2018coded,ibrahim2018coded} considered partitioning of the cache memory among the layers and proposed  algorithms for cache assignment among layers.

{\bf Our contribution:} We proposed a multipoint multicast system for on-demand video content wireless delivery 
based on coded caching and lossy successive-refinement source coding. 
The proposed system is an extension of our previous works in \cite{bayat2018coded, 2018arXiv180305702B}, where  
the coded caching message is MDS encoded and the MDS-encoded  symbols are transmitted simultaneously from multiple wireless 
{\em Edge Nodes} (ENs).  Each user can choose to receive data from up to $L$ ENs. 
The source files are encoded into two layers, a fundamental layer and a refinement layer.  Depending on the channel quality, each user may be able to decode only the fundamental layer, or both layers. In particular, using MDS codes of different rates for each layer, the successful decoding of 
a layer depends on the number of ENs that each user is able to decode. We use stochastic geometry to analyze the described system and consider 
both small and large scale fading. Our aim consists of minimizing the content delivery time subject to a target average per-user
distortion requirement, where the optimization is with respect to the cache memory allocation of two layers.
The proposed system achieves a distortion-delivery time tradeoff that is independent of the coverage area, in the limit of large area and
given EN and users density. In this sense, we say that the system is spatially scalable.


\section{system model and definition}

\label{sec:SystemModel}

We consider a wireless network with $N_E$ ENs with single antenna and $K$ users equipped with an antenna array of size $n_r$ and  $M$ units cache memory. 
All ENs are connected to the server through an error-free backhaul. The server has access to  a content library with $N$ files $\Fc = \{W_1,W_2,..., W_N\}$, each of which consists of $F$  i.i.d samples $ \sim \Nc(0,\sigma^2)$.  ENs are spatially distributed on the plane according to a two-dimensional homogeneous Poisson Point Process (PPP) of density $\lambda$. 
Each file $W_{j},  j\in [N]$ \footnote{For an integer $n$ we let $[n] = \{1,2, \ldots, n\}$.}  is source-encoded into $I=2$ successive refinement 
layers at rates $R^{S}_{i} , 1 \leq i \leq I$ bits per source sample, 
such that $|W_{n, (i)}| = R^{S}_{i}F$ bits, where $W_{j, (i)}$ denotes the $i$-th layer of the $j$-th file . 
The users assign $M_i , i\in[I]$ units out of the total $M$ memory units to cache segments (or ``subfiles'') of layer $i$.
Given a cache memory partition, we apply a separate Maddah-Ali and Niesen (MAN) coded caching schemes for each refinement layer
\cite{maddah2014fundamental}.  We let $t_i = M_i K/N, i \in [I]$, and assume that $t_i$ are integers for each $i$. The case of non-integer ratios
can be handled by memory sharing between integer points along the lines described in \cite{maddah2014fundamental}.
The MAN scheme consists of two phases: a pre-fetching and delivery phase. In the pre-fetching phase, each layer of all files ($W_{j, (i)}$) is partitioned into subfiles 
$\{W_{j,(i)}^T : T \subseteq [K], |T| = t \}$ and user $k \in[K]$ caches a subset $T$ of subfiles of layer $i$ 
if and only if $k \in T$. In the delivery phase, for a given user demand vector $\dv \in [N]^K$, for each layer $i$
the server computes the codewords $X_i$ by concatenating the blocks $X_{S}^{(i)}$ for all subsets $S \subseteq [K], |S| = t+1 $, 
where each such block is given by
\begin{equation*}  \label{ziominchia}
	X_S^{(i)} = \bigoplus_{k \in S} W_{d_k, (i)}^{S\setminus\{k\}},
\end{equation*}
The overall transmission length for layer $i$ is given by 
\begin{equation*} \label{eq:lengthOfMM}
	\Lc(X^{(i)}_S) =  \frac{F R^S_i}{{K \choose t_i}} {K  \choose t_i + 1}  =  \frac{F R^S_i K (1- \mu_i)}{1 + K \mu_i},
\end{equation*}
where $\mu_i = M_i/N$ is the fractional cache memory assigned to layer $i$.
In the proposed system, the codeword $X_i$ is divided into $L_i$ equally sized blocks, for some integer $L_1 <  L_2 \leq n_r$, 
to which  $N_E - L_i$ parity blocks are appended by using {\em Maximum Distance Separable} (MDS) coding. 
The overall transmission length length in bits of each MDS-coded blockfor layer $i$ is $\frac{\Lc(X^{(i)}_S)}{L_i}$. 
The resulting MDS-coded blocks of layers 1 and 2 are sent separately to the $N_E$ ENs, such that each EN transmits two 
distinct MDS-coded blocks for layers. The MDS-coded blocks of layer 1 and 2 are transmitted in sequence using time-division, on the wireless channel.
The transmission of all ENs is simultaneous, on the same frequency band, as in current eMBMS systems \cite{lecompte2012evolved}.
Each user $k$ is able to reconstruct the entire codeword $X_i$ if it can decode at least $L_i$ messages from $L_i$ distinct ENs. 
Depending on its location with respect to the ENs, a user may decode $\ell <L_1$ or  $L_1 \leq \ell <L_2$ or $L_2 \leq \ell$ ENs. This corresponds to 
retrieving no layers, only the fundamental layer (layer 1), or both the fundamental and the successive refinement layers (layer 1 and layer 2).
Each of these decoding outcomes yield a distortion level, and we shall consider the average per-user distortion.

\section{System Analysis}
\label{sec:SystemAnalysis}

In the following section we define the receiver model and derive the achievable ergodic rate similar to our previous 
works \cite{bayat2018coded, 2018arXiv180305702B}.  A typical user indexed by $k$ is considered at origin location  of a plane. The ENs are sorted with respect to their distance from user $k$. We assume a block fading channel with fading coherence block spanning $n$ channel uses (symbols).
The space-time signal  $\Ym_k \in \CC^{n_r \times n}$ received by user $k$ corresponding to a generic fading  block is given by
\begin{equation*} \label{eq:rec_sig}
\Ym_k = \sum_{j=1}^{N_E}  \sqrt{\beta r_{k,j}^{-\eta}}   \bfh_{k,j} \underline{\xv}_j + \Nm_k,
\end{equation*}
where $\bfh_{k,j} \in \CC^{n_r \times 1}$ is the channel vector containing the small-scale fading coefficients from EN $j$ to the antenna array of user $k$,  
$\underline{\xv}_j \in \CC^{1 \times n}$ is the coded-modulation block of the symbol sent by EN $j$, 
$r_{k,j}$ is the distance between EN $j$ and user $k$, with $\eta$ and $\beta$ being the pathloss exponent and intercept, respectively. The ENs transmit at a constant average power $P = \frac{1}{n} \EE[ \underline{\xv}_j  \underline{\xv}^\herm_j ].$
The noise samples in the matrix $\Nm_k$ are independent and identically distributed (i.i.d.) $\sim \Cc\Nc(0,N_0)$ and ${\bfh}_{k,j}$  have components 
i.i.d. $\sim \mathcal{CN}(0,1)$.
Our system utilizes the \textit{Partial Zero-Forcing} (PZF) receiver strategy, which consists of applying linear zero-forcing only with respect to the signals of the $L_2$ nearest ENs. We denote the channel matrix of the coefficients between the ENs and the user $k$'s antenna array 
by $\Hm_k = [\bfh_{k,1}, \ldots, \bfh_{k,N_E}] \in \CC^{n_r \times N_E}$, and denote by 
$\Hm_{k, \Bc}$ the submatrix formed by the columns with indices $j \in \Bc \subseteq [1:N_E]$.
Then, the PZF receiver matrix is the column-normalized version of the pseudo-inverse
\begin{equation*}
\Hm_{k,1:L_2}\ssymbol{2} = \Hm_{k,1:L_2} ( \Hm^\herm_{k,1:L_2}\Hm_{k,1:L_2})^{-1}, \label{eq:psuedo_invers}
\end{equation*}
of the channel submatrix corresponding to the closest $L_2$ ENs to user $k$.
Our scenario takes place in an infinitely extended network with an asymptotically large number of edge nodes and a high-SNR interference 
limited performance, i.e. $N_E \rightarrow \infty$ and $N_0 \rightarrow 0$. The Signal-to-Interference Ratio (SIR) at $\ell$-th stream for user $k$ is given by 
\begin{eqnarray*} \label{sir-given-phi}
\SIR_{k,\ell} &=& \frac{ r_{k,\ell}^{-\eta}  \lVert \bfh_{k,\ell}\ssymbol{2} \rVert^{-2}}{\sum_{j= L_2+1}^{\infty} r_{k,j}^{-\eta}   | \tilde{h}_{k,j} |^2 },
\end{eqnarray*}
where  $\bfh_{k,\ell}\ssymbol{2}$ denotes the $\ell$-th column of $\Hm_{k,1:L_2}\ssymbol{2}$, $\tilde{h}_{k,j} \sim \mathcal{CN}(0,1)$ and
$\lVert \bfh_{k,l}\ssymbol{2} \rVert^{-2} \sim \mathcal{X}_{2(n_r-L_2+1)}$. By considering the above $\SIR$, 
the ergodic achievable rate is defined as following
\begin{eqnarray*}
{C}_{k,\ell}(\Phi) = \EE \left [  \log \left (1+   \SIR_{k,\ell}  \right )|\Phi \right ],
\end{eqnarray*}
where $\Phi$ is the ensemble of the locations of ENs. Using the same approach of \cite{bayat2018coded}, consisting of applying Jensen's inequality 
and and replacing the terms in the SIR denominator by their ensemble average, we obtain a  quasi-lower bound on the 
ergodic achievable rate of user $k$ as
 \begin{equation} \label{eq:Clb}
 {C}_{k,\ell}^{\rm qlb}(\tilde{\rho}) = \EE \left[ \left . \log \left(  1+  \tilde{ \rho}_{k,\ell} \mathcal{X}_{2(n_r-L_2+1)} \right ) \right | \Phi \right],
 \end{equation}
 where we define the approximated conditional {local-average SIR} as 
 \begin{align}  \label{eq:local_sir}
 \tilde{\rho}_{k,\ell} & =  \frac{r_{k,L_2}^{\eta-2}}{r_{k,\ell}^{\eta} } \frac{\eta-2}{2 \pi \lambda}  \quad 1\leq \ell \leq L_2 .
 \end{align}
Notice that in the rate expression (\ref{eq:Clb}), the expectation is taken with respect to the small-scale fading, 
but it is conditional  to the placement of the NEs  and user $k$  (system geometry). 
This corresponds to separating the time scale of the small-scale fading from the time scale of the geometry variation, due to mobility.
 The expectation with respect to $\mathcal{X}_{2(n_r-L_2+1)}$ in  (\ref{eq:Clb})
 can be calculated in closed form as
 \begin{equation}  \label{eq:Clb-explicit}
 {C}_{k,\ell}^{\rm qlb}(\tilde{\rho}_{k,\ell})
 = \EE \left[ \left . \log \left( 1+  \tilde{\rho} _{k,\ell} \mathcal{X}_{n_r-L_2+1}\right) \right | \Phi \right] 
 = \mathcal{I}_{(n_r-L_2+1)} ( \tilde{\rho}  _{k,\ell})\log (e),
 \end{equation}
where $ \mathcal{I}_M(\mu)$ is given by 
 \begin{equation*}
 \label{eq:exp_int}
 \mathcal{I}_M(\mu)  =  \; \Pi_M(-1/\mu)E_i(1,1/\mu) 
   + \sum_{m=1}^{M-1}\frac{1}{m} \Pi_m(1/\mu) \Pi_{M-m}(-1/\mu)
 \end{equation*}
 with $ \Pi_n(x) = e^{-x}\sum_{i=0}^{n-1}\frac{x^i}{i!}$ and with the exponential integral function defined as
 $E_i(n,x) = \int_1^\infty t^{-n} e^{-xt} dt$.
The  $\ell$-th EN transmits the modulated blocks for layer $i$ with PHY rate $R^{C}_{i}$. 
The probability of decoding error for the $i$-th layer is given by 
\begin{equation*}
P({C}_{k,L_i}^{\rm qlb}(\tilde{\rho}_{k,L_i}) \leq R^{C}_{i} ), 
\end{equation*}
where the probability is with respect to the joint distribution of the EN distances from user $k$.
The MSE distortion function for a Gaussian source  with source rate coding ($R^{S}_i$) is given by 
\begin{equation}
D_i= \sigma^2 2^{-2R^{S}_i}, 1\leq i\leq I.
\end{equation}
We define $\sigma^2 = D_{\max} = 1$ and write the average distortion for user $k$ as
\begin{align}\label{eq:Dis_def}
& D_k = \PP(C^{\rm qlb}_{k,L_1}(\tilde{\rho}_{k,L_1}) \leq R^C_i)  + \PP(R^{C}_1 < C^{\rm qlb}_{k,L_1}(\tilde{\rho}_{k,L_1}) , 
C^{\rm qlb}_{k,L_2}(\tilde{\rho}_{k,L_2}) \leq R^{C}_2)  2^{-2 R^{S}_1} \\ &
+ \PP(R^{C}_1< C^{\rm qlb}_{k,L_1}(\tilde{\rho}_{k,L_1}) , R^{C}_2 < C^{\rm qlb}_{k,L_2}(\tilde{\rho}_{k,L_2}))2^{-2( R^{S}_1+R^{S}_2)}. \nonumber
\end{align}
The time necessary to deliver all requested files depends on the overall transmission time for each layer. The delivery latency is given by
\begin{equation}  \label{latency}
T = \sum_{i=1}^{I} \frac{\Lc(X_i)}{wR_i^C}= \sum_{i=1}^{I}\frac{FR^S_i}{w} \frac{1}{R^{C}_{i} L_i} \frac{K(1-\mu_i)}{1+K \mu_i},
\end{equation}
where $w$ is the bandwidth of the wireless channel and $R_i^S$ and $R_i^C$ are source coding and channel coding rate for layer $i$.
Expressions (\ref{eq:Dis_def}) and (\ref{latency}) establish an achievable tradeoff between average distortion and delivery time, that we wish to optimize with respect to the cache allocation parameters $\muv = \{\mu_i\}$, the channel coding rates $\{R^C_i\}$, and the MDS ``macro-diversity orders'' $\{L_i\}$.  
For given cache allocation $\bfmath{\mu}$  
the constraint on cache allocation among layers is given as
\begin{equation*}
\sum_{i=1}^{I} M_i F R^{S}_{i} = \mu N F \sum_{i=1}^{I} R^{S}_{i}.
\end{equation*}
where $\mu$ denotes the fraction of total source-encoded library bits cached at each user.
Eliminating $F$ from both sides, dividing by $N$, and dividing by $\sum_{i=1}^{I}R^{S}_{i}$, we obtain the cache allocation constraint as
\begin{equation*}
\sum_{i=1}^{I}  \mu_i \hat{R}_i^S = \mu
\end{equation*}
where we define $ \hat{R}_i^S := \frac{R^{S}_{i}}{\sum_{i=1}^{I} R^{S}_{i}}$ 
We assume that the source coding rates are given (e.g., standard and high definition of a given video format). Hence, 
our objective consists of minimizing the delivery latency subject to a target average distortion constraint $D_0$. This yields the following problem:
 \begin{equation}\label{eq:opt_main}
 \begin{aligned}
 &~ \underset{R^{C}_{1}, R^{C}_{2},\mu_1, \mu_2}{\text{minimize}}
 & & T \\
 & ~\text{subject to}
 & & D_k \leq D_0 \\
 &~ &&  \sum_{i=1}^{I}  \mu_i \hat{R}_i^S = \mu .
 \end{aligned}
 \end{equation}

\subsection{Multidimensional decoding error probability  }

Given a homogeneous PPP $\Phi$ of density of $\lambda$,  let $r_n$ denote the 
$n$-th shortest distance of points of $\Phi$ from the origin.
Then, the Probability Density Function (PDF) of $r_n$ is given by \cite{martin2014multi}
\begin{equation*}
f_{ r_{n}}(v) = \frac{2(\pi \lambda)^{n}}{(n-1)! } v^{2n-1} e^{-\pi \lambda v^2},  \;\; v \geq 0 
\end{equation*}
and the joint PDF of $r_{\ell}$ and $r_{n}$ with $1 \leq \ell < n$ is given by
\begin{equation*}
f_{r_{\ell}, r_{n}}(u,v) = \frac{4 (\pi \lambda)^{n}}{(n-\ell-1)! \, (\ell - 1)!}(v^2-u^2)^{n-\ell-1} v u^{2\ell-1} e^{-\pi \lambda v^2}, \;\;\; u,v \; \geq 0.
  \label{jointrr}
\end{equation*}
The conditional PDF of $r_{\ell}$ on $r_{k}$ with $1 \leq \ell < k$ and $ v \; \geq u \; \geq 0$ is given by
\begin{align*}
f_{r_{\ell}\mid r_{k}}(u \mid v)  =   
 \sum\limits_{n=0}^{k - \ell -1} \alpha_{n, \ell, k}  (-1)^n v^{-2(n+\ell)} u^{2(n+\ell)-1}  \;\;\;  
\end{align*}
and when $v < u$, $f_{r_{\ell}\mid r_{k}}(u \mid v) = 0$,
where we define \[\alpha_{n, \ell, k} = \frac{2(k-1)!  }{(k-\ell-1)! \, (\ell - 1)!} \binom{k -\ell - 1}{n}.\]
By substituting $\ell = L_2$ in (\ref{eq:local_sir}) and  considering a given $R^C_{2}$, the threshold  on $L_2$-th EN's distance  $ \widehat{r}_{k,L_2}> r_{k,L_2}$  such that $R^{C}_{2}<\ {C}_{k, L_2}^{\rm qlb}$, is given by
\begin{equation*} \label{eq"given_r_2}
\widehat{r}_{L_2} := \sqrt{\frac{\eta-2}{2 \pi \lambda\left (  {C}_{k, L_2}^{\rm qlb} \right )^{-1} (R^{C}_{2})}},
\end{equation*}	 
where $\left (  {C}_{k, L_2}^{\rm qlb} \right )^{-1} (R^{C}_{2})$ is the inverse of the function in (\ref{eq:Clb-explicit}).
By substituting $\ell = L_1$ and $r_{k,L_2} = v$ in (\ref{eq:local_sir}) and  considering a given $R^C_{1}$, 
the threshold  on $L_1$-th EN's distance  $ \widehat{r}_{k,L_1}> r_{k,L_1}$  such that $R^{C}_{1}<\ {C}_{k, L_1}^{\rm qlb}$, is given by
\begin{equation*}
\widehat{r}_{k,L_1} := \left ( {\frac{\eta-2}{2 \pi \lambda }\frac{v^{\eta-2}}{\left ( {C}_{k, L_1}^{\rm qlb} \right )^{-1} (R^{C}_{1})}} \right)^{1/{\eta}}.
\end{equation*}	
The successful decoding probability of layer one conditioned on $r_{k,L_2} = v$ is given by
\begin{equation*}
\PP(R^{C}_1<\ {C}_{k, L_1}^{\rm qlb} \mid r_{k,L_2}=v) =\int_{0}^{\min (v, \widehat{r}_{k,L_1})} f_{r_{k,L_1}\mid r_{k, L_2}}(u \mid v) \mathrm{d} u.
\end{equation*}
By defining $ a := \sqrt{\frac{\eta-2}{2\pi\lambda \left ( {C}_{k, L_1}^{\rm qlb}\right )^{-1}(R^{C}_{1})}}$ such that $v \leq \widehat{r}_{k,L_1}$, $v \leq a$ are equivalent.
By considering the two cases $v \leq a$  or $v > a$, we calculate the integral as follows

\begin{equation*}
\textbf{Case 1	: } \PP(R^{C}_1<\ {C}_{k, L_1}^{\rm qlb} \mid  r_{k,L_2} = v , v \leq a)=\int_{0}^{v} f_{r_{k, L_1}\mid r_{k,L_2}}(u \mid v) \mathrm{d} u = 1, 
\end{equation*}
\begin{equation*}
\textbf{Case 2   : } \PP(R^{C}_1<\ {C}_{k, L_1}^{\rm qlb} \mid  r_{k,L_2} = v, v > a ) =\int_{0}^{\widehat{r}_{k,L_1}} f_{r_{k,L_1}\mid r_{k,L_2}}(u \mid v) \mathrm{d} u 
= \sum\limits_{n=0}^{L_2 - L_1 -1} \alpha'_{n,L_1,L_2}  \left( \gamma_{{1}} \right)^{\eta'} v^{-\eta'}, 
\end{equation*}
where $\eta' = \frac{2(n+L_1)}{\eta}$, $\gamma_{{1}}= \frac{\eta-2}{2\left ( {C}_{k,L_1}^{\rm qlb}\right )^{-1}(R^{C}_{1})}  $ and
$\alpha'_{n,L_1,L_2} =  \frac{(L_2-1)!  (-1)^n \binom{L_2-L_1- 1}{n} }{(L_2-L_1-1)! \, (L_1- 1)!\, (n+L_1)} .$ 

By using these two conditional integrals, the successful decoding probability of the first layer can be derived as follows
\begin{align*}
\PP(R^{C}_1<\ {C}_{k, L_1}^{\rm qlb} )
&=  \int_{0}^{a} 	\PP( R^{C}_1<\ {C}_{k, L_1}^{\rm qlb}\mid r_{k,L_2}=v , v<a) f_{r_{k,L_2}}(v) \mathrm{d} v \nonumber \\
&\quad+  \int_{a}^{\infty} 	\PP( R^{C}_1<\ {C}_{k, L_1}^{\rm qlb} \mid r_{k,L_2}=v , v>a) f_{r_{k, L_2}}(v) \mathrm{d} v  \nonumber\\
&= 	\frac{1}{\Gamma(L_2)} \bar{\Gamma} \left(L_2, \gamma_{{1}} \right)+\sum\limits_{n=0}^{L_2 - L_1-1} \alpha'_{n,L_1,L_2} \left( \gamma_{{1}} \right)^{\eta'}\Gamma \left(L_2-\eta', \gamma_{{1}} \right), \nonumber
\end{align*}	
where  ${\Gamma}(s, x) $ and $\bar{\Gamma}(s, x) $ are upper and lower incomplete gamma functions, respectively.
By conditioning the joint probability on $r_{k,L_2}$, the conditional joint probability is given by
\begin{align*}\label{eq:condJoint1}
\PP(R^C_1 \leq {C}_{k, L_1}^{\rm qlb} , R^C_2> {C}_{k, L_2}^{\rm qlb} \mid r_{k,L_2} = v)
& = \PP(R^C_2> {C}_{k, L_2}^{\rm qlb} \mid r_{L_2})\PP(R^C_1 \leq {C}_{k, L_1}^{\rm qlb} \mid r_{k,L_2} = v) \nonumber\\
& =  \mathbb{1} \left(r_{L_2} > \widehat{r}_{L_2}\right) \PP(R^C_1 \leq {C}_{k, L_1}^{\rm qlb}\mid r_{k,L_2} = v).
\end{align*}

By considering the two cases  $a \leq \widehat{r}_{k,L_2}$ or $a > \widehat{r}_{k,L_2}$ the joint probability is calculated as following.
The expression $a \leq \widehat{r}_{k,L_2}$ is equivalent to  $\left ( {C}_{k, L_1}^{\rm qlb}\right )^{-1}(R_{C_1}) \geq \left ({C}_{k, L_2}^{\rm qlb}\right )^{-1}(R_{C_2}))$. Notice that ${C}_{k,\ell}^{\rm qlb}(\tilde{\rho} _{k,\ell})$ is strictly monotonically increasing in $\tilde{\rho} _{k,\ell}$, and this simplify  the expression $\left ( {C}_{k, L_1}^{\rm qlb}\right )^{-1}(R^C_1) \geq \left ( {C}_{k, L_2}^{\rm qlb}\right )^{-1}(R_2^C)$ to $R^C_1 \geq R^C_2$.

\begin{equation*}
\textbf{ Case 1	: }  \PP(R^C_1 \leq {C}_{k, L_1}^{\rm qlb}, R^C_2 > {C}_{k, L_2}^{\rm qlb} \mid R^C_1 \geq R^C_2) 
= \sum\limits_{n=0}^{L_2 - L_1-1} \alpha'_{n,L_1,L_2}  \left( \gamma_{1} \right)^{\eta'}\Gamma \left(L_2-\eta', \gamma_{2}  \right)
\end{equation*}	
\begin{align*}
 \textbf{Case 2	: } \PP(R^C_1 \leq {C}_{k, L_1}^{\rm qlb}, R^C_2 > {C}_{k, L_2}^{\rm qlb} \mid R^C_1 < R^C_2) 
 = & 	\frac{1}{\Gamma(L_2)} \bar{\Gamma} \left(L_2, \gamma_{1} \right)- \frac{1}{\Gamma (L_2)} \bar{\Gamma} \left( L_2,  \gamma_{2}  \right) \nonumber \\
& + \sum\limits_{n=0}^{L_2 - L_1-1} \alpha'_{n,L_1,L_2}  \left( \gamma_{1} \right)^{\eta'}\Gamma \left(L_2-\eta', \gamma_{1}  \right),
\end{align*}
where $\gamma_{{2}}= \frac{\eta-2}{2\left ( {C}_{k,L_2}^{\rm qlb}\right )^{-1}(R^{C}_{2})}  $.
Similarly, the last joint probability is given by two cases as follows
\begin{align*}
\textbf{Case 1	: } \PP(R^C_1  \leq {C}_{k, L_1}^{\rm qlb}, R^C_2 \leq {C}_{k, L_2}^{\rm qlb} \mid R^C_1 \geq R^C_2)  
= & \frac{1}{\Gamma (L_2)}\bar{\Gamma} \left( L_2, \gamma_{1} \right)
- \sum\limits_{n=0}^{L_2 - L_1-1} \alpha'_{n,L_1,L_2}  \left( \gamma_{1} \right)^{\eta'}\Gamma \left(L_2-\eta', \gamma_{2}  \right) \nonumber  \\
&+	 \sum\limits_{n=0}^{L_2 - L_1-1} \alpha'_{n,L_1,L_2}  \left( \gamma_{1} \right)^{\eta'}\Gamma \left(L_2-\eta', \gamma_{1}  \right)
\end{align*}
\begin{equation*}
\textbf{Case 2	: }\PP(R^C_1 \leq {C}_{k, L_1}^{\rm qlb}, R^C_2 \leq {C}_{k, L_2}^{\rm qlb} \mid R^C_1 < R^C_2) 
=  \frac{1}{\Gamma (L_2)}\bar{\Gamma} \left( L_2,  \gamma_{{2}}  \right).
\end{equation*}

\section{sub-optimal rate and cache allocation}
\label{sec:StatementOfOptimization}

In this section, we provide a feasible but generally sub-optimal  solution to the
non-convex optimization problem defined in (\ref{eq:opt_main}). 
This optimization has two constrains for cache allocation and minimum distortion requirement. 
Since the problem is non-convex and does not seem to have some especially appealing structure 
that can be exploited for its efficient solution, mainly due to the complicated dependency of the probabilities of layer decoding error on the channel coding 
rates, we propose an iterative method described in Algorithm \ref{alg:search}.
We denote by $T( R^{C}_{1},  R^{C}_{2},\mu_1,\mu_2)$ the value of $T$ in \eqref{eq:Dis_def} for given  
$R^{C}_{1},  R^{C}_{2},\mu_1,\mu_2$. Then, the algorithm applies alternate minimization
by fixing $\mu_1,\mu_2$ and minimizing with respect to $R^{C}_{1},  R^{C}_{2}$, and for the found values of
$R^{C}_{1},  R^{C}_{2}$ minimizing with respect to $\mu_1,\mu_2$. 

We can handle the constrained minimization of $T( R^{C}_{1},  R^{C}_{2},\mu_1,\mu_2)$ subject to the distortion constraint with respect to $R^{C}_{1},  R^{C}_{2}$ 
for fixed $\mu_1,\mu_2$ using the \textit{Particle Swarm Optimization} (PSO) algorithm \cite{parsopoulos2002particle}. 
PSO performs a heuristic search in order to find a good feasible point. It is useful for minimizing a function with linear/nonlinear  
inequality constraints and it uses a penalty function technique to solve an unconstrained optimization instead of its constrained counterpart. 
Interestingly, the solution of the minimization with respect to $\mu_1,\mu_2$ for fixed $R^{C}_{1},  R^{C}_{2}$ can be found in closed form. 
Notice that the constraint $D_k \leq D_0$ involves only the variables $R^{C}_{1},  R^{C}_{2}$, therefore, it does not play any role in the optimization of
$\mu_1,\mu_2$.  
The function  $T( R^{C}_{1},  R^{C}_{2},\mu_1,\mu_2)$  is convex in the vector $\muv = \left[\mu_1, \mu_2\right]$. 
 The minimization 
with respect to $\mu_1,\mu_2$ at the $t$-th step of our iterative algorithm is given by:
\begin{equation*}\label{eq:opt_man}
\begin{aligned}
&~ \underset{\mu_1, \mu_2}{\text{minimize}}
& &T(R^{C}_{1,(t)},R^{C}_{2,(t)},\mu_1,\mu_2)  \\
& ~\text{subject to}
& &   \sum_{i=1}^{2}  \mu_i \hat{R}_i^S = \mu  
\end{aligned}
\end{equation*}
The partial Lagrangian function (not taking into account the non-negativity constraints) is given by 
\begin{equation*} \label{eq:sub-opt_lg}
\Lc (\lambda , {\bfmath{\mu}})= \sum_{i=1}^{2} \frac{R^{S}_{i}}{R^{C}_{i,(t)}}\frac{K(1-\mu_i)}{L_i(1+K \mu_i)} +\lambda ( \sum_{i=1}^{2}  \mu_i \hat{R}_i^S - \mu ).
\end{equation*}

After some algebra, we have  sub-optimal solution. 
The value for $\mu_1, \mu_2$ for the next iteration are given by 
\begin{align*}
\label{eq:opt_mu}
 \mu_{i,(t+1)}= \frac{1}{K} \left (  \frac{(1+K\mu) \sqrt{\alpha_i /\hat{R}_i^S}}{\sum_{j=1}^{I} \sqrt{\alpha_j  \hat{R}_j^S  }} -1 \right ).
\end{align*}

\begin{algorithm}
	\caption{Searching strategy }
	\label{alg:search}
	\begin{algorithmic}[1]
		\Require{ $\mu$, $K$, $\eta$, $n$ and  $L_i$, $R^S_i,\, i\in[2]$ :} 
		\State Set the initial value $\mu_{1,{(1)}} =\mu$ and $\mu_{2,{(1)}}=\mu$
		\For{$t= 1:n$}
		\State  $[ R^{C}_{1,(t)}, R^{C}_{2,(t)} ]= \argmin T( R^{C}_{1,(t)},  R^{C}_{2,(t)},\mu_{1,(t)},\mu_{2,(t)}) $ 
		\State Set $\alpha_{i} =\frac{R^S_i}{R^{C}_{i,(t)} L_i},\, i\in [2]$ 
		\State Set the $\mu_{i,{(t+1)} }= \frac{1}{K} \left (  \frac{(1+K\mu) \sqrt{\alpha_i /\hat{R}_i^S}}{\sum_{j=1}^{2} \sqrt{\alpha_j \hat{R}_j^S}} -1 \right ),\,i\in [2]$
		\EndFor
	\end{algorithmic}
\end{algorithm}

\section{results and discussions}
\label{sec:NumericalResults}

\begin{figure}[t]
	\centering
	\includegraphics[width=0.65\textwidth]{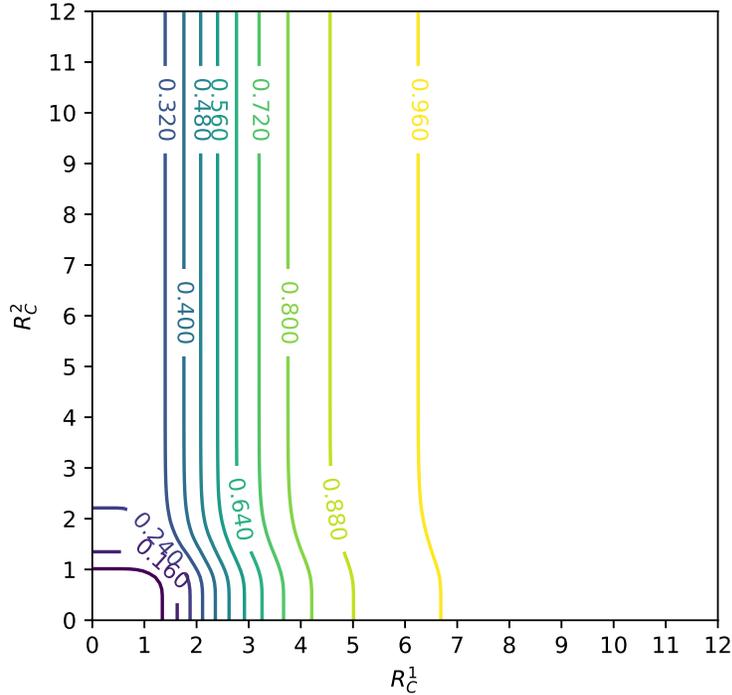}
	\caption{ Level sets of distortion versus channel coding rates }
	\label{fig:levelset}
\end{figure}

In this section we provide numerical examples to illustrate the sub-optimal
solution for the delivery latency in (\ref{eq:opt_main}). 
We considered realistic values of the pathloss exponent $\eta = 3.75$ and the number of antennas $n_r=8$ at the user receivers and source coding rate $R^S_1 =1$ and $R^S_2=2$. The macro diversity for the receiving layers are considered to be $L_1=2$ and $L_2=4$.
The level set for the average distortion function as defined in (\ref{eq:Dis_def}) is plotted in Fig. \ref{fig:levelset}. It should be noted that the average distortion can be divided into two distinct regions. In the first region, where $ 0 \leq D_k \leq 2^{-2R_{1}^S}$ the level sets are bounded such that the channel coding rate for both layers is below the boundaries. In the second region with $ 2^{-2R_{1}^S}< D_k \leq 1$, the rate $R^C_{2}$ goes to the boundaries with the second layer having a very high probability of being incorrectly decoded. 

\begin{figure}[t]
	\centering
	\includegraphics[width=0.7\textwidth]{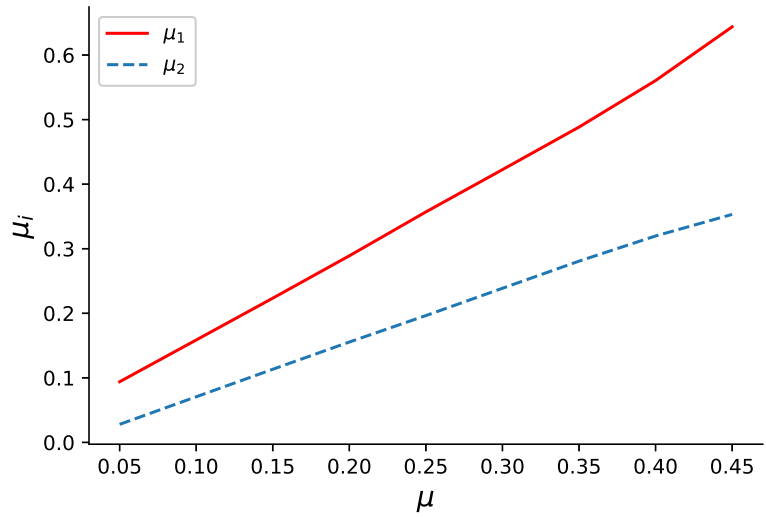}
	\caption{ Cache allocation among layers }
	\label{fig:mu}
\end{figure}

\begin{figure}[t]
	\centering
	\includegraphics[width=0.7\textwidth]{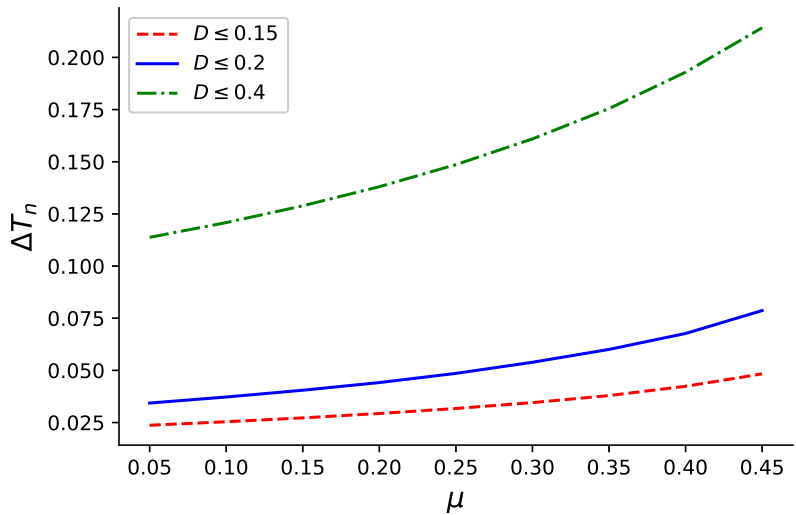}
	\caption{Improvement of delivery latency of optimized cache allocation over uniform cache allocation}
	\label{fig:delay}
\end{figure}

In Fig. \ref{fig:mu}, the cache allocation among layers is illustrated for target distortion  $D_k\leq 0.2$. 
 We compare the performance of the cache allocation according to the solution of the optimization in (\ref{eq:opt_main}) with the scenario where the cache allocation is given by $\mu_1 = \mu_2= \mu$. The metric for this comparison takes their delivery latency into account and is given by 
\begin{equation*}
 \Delta T_n = \frac{T_{\rm unif}-T_{\rm opt}}{T_{\rm opt}}
\end{equation*}
where $T_{\rm opt}$ and $T_{\rm unif}$ are the average delivery latencies for the former and latter scenario, respectively. 
The comparison is illustrated in Fig. \ref{fig:delay} for various distortion requirements. The sub-optimal iterative method with heuristic searching decreases the delay between $2-20 \%$. As can be seen in Fig. \ref{fig:delay}, the proposed optimization achieves significantly better performance  for high target distortion values.

\section{Conclusion}

In this paper, we studied  an extension of the classical coded caching scheme with the goal of
achieving spatial scalability and graceful degradation. Spatial scalability is obtained by sending the coded caching codewords
simultaneously from multiple Edge Nodes, using MDS coding in order to enforce a desired level of macro-diversity. 
Graceful degradation is obtained by using successive refinement source coding and encoding the source files into a fundamental layer 
and a refinement layer. 
The scheme can be also interpreted as the concatenation of coded caching with a multiple-description code, where the users that can decode $L_1$ descriptions achieve a distortion level $D_1$ of the fundamental layer, and the users that can decode $L_2 > L_1$ descriptions achieve a distortion level $D_2 < D_1$ of the refinement layer. 
We studied the optimization of the delivery time subject to an average per-user distortion constraint. The optimization is with respect to 
the channel coding rates for the transmission of the two layers, and the memory allocation parameter of the user caches between the two layers. 
The optimization is non-convex, but it can be handled by alternating minimization.

The proposed system could be applied to the lossy transmission of video-on-demand (unicast traffic) via multipoint multicasting, 
conceptually similar to a caching extension of eMBMS, to handle individual user demands and not only broadcast of common content (such as Live TV).
The approach can be extended to more than 2 layers, but the main bottleneck consists of the analysis of 
the probability of successful decoding of a given number $\ell$ of layers, since this involves the joint distribution of the first $L$ distances
of a PPP with respect to the origin. 
\balance

{\small
\bibliographystyle{IEEEtran}
\bibliography{references}
}

\end{document}

%% file: symbols.tex
\usepackage{mathbbol}

\def\mindex#1{\index{#1}}

\def\sq{\hbox{\rlap{$\sqcap$}$\sqcup$}}
\def\qed{\ifmmode\sq\else{\unskip\nobreak\hfil
\penalty50\hskip1em\null\nobreak\hfil\sq
\parfillskip=0pt\finalhyphendemerits=0\endgraf}\fi\medskip}

\long\def\defbox#1{\framebox[.9\hsize][c]{\parbox{.85\hsize}{%
\parindent=0pt
\baselineskip=12pt plus .1pt      
\parskip=6pt plus 1.5pt minus 1pt 
 #1}}}

\long\def\beginbox#1\endbox{\subsection*{}%
\hbox{\hspace{.05\hsize}\defbox{\medskip#1\bigskip}}%
\subsection*{}}

\def\endbox{}

\newsavebox{\junk}
\savebox{\junk}[1.6mm]{\hbox{$|\!|\!|$}}

\def\argmin{\mathop{\rm arg\, min}}

\def\bfh{{\bf h}}

\def\bfmath#1{{\mathchoice{\mbox{\boldmath$#1$}}%
{\mbox{\boldmath$#1$}}%
{\mbox{\boldmath$\scriptstyle#1$}}%
{\mbox{\boldmath$\scriptscriptstyle#1$}}}}

\def\bfmY{\bfmath{Y}}

\def\bfmhhaY{\bfmath{\hhaY}} 
\def\bfmhhaY{\hbox to 0pt{$\widehat{\bfmY}$\hss}\widehat{\phantom{\raise 1.25pt\hbox{$\bfmY$}}}}

\def\til={{\widetilde =}}

 \def\FRAC#1#2#3{\genfrac{}{}{}{#1}{#2}{#3}}

\def\ddtp{{\mathchoice{\FRAC{1}{d^{\hbox to 2pt{\rm\tiny +\hss}}}{dt}}%
{\FRAC{1}{d^{\hbox to 2pt{\rm\tiny +\hss}}}{dt}}%
{\FRAC{3}{d^{\hbox to 2pt{\rm\tiny +\hss}}}{dt}}%
{\FRAC{3}{d^{\hbox to 2pt{\rm\tiny +\hss}}}{dt}}}}

\def\average#1,#2,{{1\over #2} \sum_{#1}^{#2}}

\def\eye(#1){{\bf(#1)}\quad}

\def\eq#1/{(\ref{e:#1})}

\newcommand{\beqn}[1]{\notes{#1}%
\begin{eqnarray} \elabel{#1}}

\newcommand{\eeqn}{\end{eqnarray} }

\newcommand{\beq}[1]{\notes{#1}%
\begin{equation}\elabel{#1}}

\newcommand{\eeq}{\end{equation}}

\def\bdes{\begin{description}}
\def\edes{\end{description}}

\newcounter{rmnum}

\newcounter{anum}

{\end{list}}

\def\ass(#1:#2){(#1\ref{#1:#2})}

\def\ritem#1{
\item[{\sf \ass(\current_model:#1)}]
}

\newenvironment{recall-ass}[1]{%
\begin{description}
\def\current_model{#1}}{
\end{description}
}



%% file: macros.tex
\newfont{\bb}{msbm10 scaled 1100}
\newcommand{\CC}{\mbox{\bb C}}
\newcommand{\PP}{\mbox{\bb P}}

\newcommand{\EE}{\mbox{\bb E}}

\newcommand{\dv}{{\bf d}}

\newcommand{\xv}{{\bf x}}

\newcommand{\Hm}{{\bf H}}

\newcommand{\Nm}{{\bf N}}

\newcommand{\Ym}{{\bf Y}}

\newcommand{\Bc}{{\cal B}}
\newcommand{\Cc}{{\cal C}}

\newcommand{\Fc}{{\cal F}}

\newcommand{\Lc}{{\cal L}}

\newcommand{\Nc}{{\cal N}}

\newcommand{\muv}{\hbox{\boldmath$\mu$}}

\newcommand{\herm}{{\sf H}}

\newcommand{\SIR}{{\sf SIR}}

\usepackage{hyperref}
\hypersetup{
    bookmarks=true,         
    unicode=false,          
    pdftoolbar=true,        
    pdfmenubar=true,        
    pdffitwindow=false,     
    pdfstartview={FitH},    
    pdfnewwindow=true,      
    colorlinks=true,       
    linkcolor=red,          
    citecolor=green,        
    filecolor=blue,      
    urlcolor=blue           
}